# Enhancing Patient Appointments Scheduling that Uses Mobile Technology

Godphrey G. Kyambille

Computation and Communication Science &Engineering
Nelson Mandela African Institution of Science and Tech
Arusha, Tanzania

Khamisi Kalegele

Computation and Communication Science &Engineering
Nelson Mandela African Institution of Science and Tech
Arusha, Tanzania

*Abstract—* **Appointment scheduling systems are utilized mainly by specialty care clinics to manage access to service providers as well as by hospitals to schedule patient appointments. When attending hospitals in Tanzania, patients experience challenges to see an appropriate specialist doctor because of service interval inconsistency. Timely availability of doctors is critical whenever a patient needs to see a specialist doctor for treatment and a serious bottleneck lies in the application of appropriate technology techniques to enhance appointment scheduling. In this paper, we present a mobile based application scheduling system for managing patient appointments. Furthermore, forthcoming opportunities for the innovative use of the mobile based application scheduling system are identified.**

*Keywordst Mobile application, Hospital, Appointment scheduling, Patient*

## I. INTRODUCTION

An effective patient appointment scheduling system is very critical in hospitals to ensure effective and efficient service delivery in the health sector in Tanzania. Yet in order to target efficient appointment scheduling, there is a need for appropriate management and quality evaluation of the scheduling system. Most patients complain about the time spent between walking into the hospital and being attended by hospital staff, especially doctors. And this calls for proper handling. The proposed mobile application for patient appointment scheduling is poised to effectively facilitate delivery of health services in Tanzanian hospitals.

Making appointments over the mobile phone provides more benefits. These include time saving as staff spends less time in attending patients compared to paper-based appointments where patients need to fill in lots of forms. There is no waste of time in queues when a mobile application based patient appointment scheduling system is used. Furthermore, the automated appointment reminder in the mobile appointment scheduling system also saves time as hospital operators won't be required to call and send SMS to patients reminding them of their appointments. Mobile phone based appointment system allows for 24 hours convenient scheduling and patients can make appointments at any time compared to making appointments physically by showing up at hospitals, which can be done only during working hours.

The World Health Organization (WHO) conducted a global survey in 2011 involving 114 nations and found that mobile devices are used in almost all countries but they vary on the uptake level: some use the mobile devices to send reminders to patients by sending text messages on their appointment, telemedicine, accessing patient records, monitoring patients and symptoms diagnosis [1]

There is an emphasis on the need to change in the way hospital services are offered by adapting e-Health technologies in order to achieve the national vision of applying information and communication technologies (ICT) in the health sector [2]. In one study, it was reported that waiting time for patients who attended their disability hospital appointments before receiving treatment was reduced due to enhancement of the system they implemented for triage patient appointment [3]. The hospitals' use of mobile technologies in scheduling appointments can facilitate rapid response; physicians can prescribe medication more safely, and there is high possibility of improving the patient's hospital records during daily clinic visits [4].

In solving patient waiting time, a discrete simulation model was proposed to illustrate how to improve clinic performance [5]. Based on dynamic and complexity of healthcare scheduling system when applying the simulation model, results show physicians' work time when combined with patient's admission time changing would reduce patients waiting time up to 73%.

Mobile appointment systems have been recommended for use in the health sector in order to improve the workflow, and as a result enhance scheduling of patients based on their priorities [6]. The patient's use of a mobile application system in making appointment allows him/her to request for appointment, negotiate with the clinic if the appointment is urgent, and choose his/her time preference among the available time slots [7].

This study's objective is to enhance the appointment scheduling system via a mobile application, which facilitates assigning time slots to patients whenever they make appointments and prioritize patients with high precedence. Patients who forget their appointments can receive a reminder alert on the upcoming appointment, and the clinic can track appointments and health performance of their patients.

## II. OVERVIEW OF THE APPOINTMENT SCHEDULING SYSTEM

In this section, we review the literature on the use of the mobile technology in appointment scheduling by hospitals. The primary objective is to find out exactly areas where improvement can be made to support the health landscape in Tanzania.

Appointment scheduling via paper-based system requires patients to be at the hospital, fill in registration forms and return them to the registration desk, and patients are then assigned to the desired doctor. Sometimes, patients place hospital identification cards or appointment cards in the dedicated box near the doctor's room, and then wait in the queue to be called by the nurse. Cards are placed in the order of first come, first serve (FCFS), whereby the patient who came early is the first to be served and the last to show up waits on the queue. Patient information in the paper-based appointment system cannot be easily corrected when changes need to be made: another form will need to be filled in, and the data entry registration desk staff experience problems in reading information written in the paper appointment forms, and it is difficult to retrieve patient details when required as you need the entire appointment application form ([8]). This type of appointment scheduling system has a range of constraints, such as patients being required to fill in appointment forms upon arrival at the hospital, and there is no possibility to register while at home or any place as a result, patients spend a lot of time waiting in queues, are required to follow dates of appointment assigned by the registration desk, and there is no mechanism for patients notification when appointments are postponed. Additional, managing paper-based hospital appointment system is difficult to manage, hence the need for a new method. The use of mobile appointment scheduling can enhance hospital appointments as it will allow patients to make appointments before going to the hospital. Patients can be reminded of the appointment as well. The clinic can monitor patient's performance while on the provided treatment, and the patient can select desired date of appointment based on his/her wishes.

Near field communication technology is a wireless communication that is used to transmit data at a short range of distance, approximately 10cm ([9]). The intelligent agent system was developed for appointment scheduling where patients can register and make appointments through mobile devices and eliminate the registration desk staff ([10]). Smart technologies for mobile appointment have been developed where patients use mobile and Near Field Communication Technology (NFC) ([11]). Patients need to tap their NFC cards into NFC readers at the main entrance gate of the hospital, and once there is an information match, the other scheduling procedures follows.

Ingagepatient.com is an online appointment scheduling system where patients need to register or sign up online in order to make appointments. New patients are required to have email accounts at the initial stage of registration. Once registered, patients are required to fill appointment forms at their own pace without queuing.

To facilitate effective service delivery in hospitals in Tanzania, a mobile system for patient appointment is proposed where patients need to download and install the application in their mobile phones, and then they can register on the application and receive username and password which can be used for login in the application ([12]). After login, patients need to select filtration type, and a list of doctors is displayed based on the selected filter. Then, the patient is required to select a desired doctor and his/her schedule is displayed, and finally the patient can make an appointment based on the doctor's free time slot.

## III. THE PROPOSED MOBILE APPOINTMENT SCHEDULING SYSTEM

The proposed Mobile Appointment Scheduling System (MASS) aims at enhancing appointment scheduling in hospitals by allowing patients to register for appointments through mobile phones at their own time wherever they are, and make an appointment on their desired slot of time.

### A. Requirements analysis

A modified wave appointment scheduling algorithm is a proposed approach for patient appointment scheduling in which patients are scheduled in 10 minutes and more than one patient is booked toward the beginning of every hour and the hour end is left open, permitting the specialist to make up for lost time, if needed. When the patient is attended to in less than the allocated time, the remaining time will be assigned to the next patient and idle time for waiting and doctor work overload will be reduced. Patient waiting time is expected to be reduced from 3 hours to half an hour. The modified wave scheduling will facilitate patient flow and rise patient satisfaction.

In order to accomplish the patient appointment request Fig.1 summarizes the involvement of actors and their collaboration among themselves and the system.

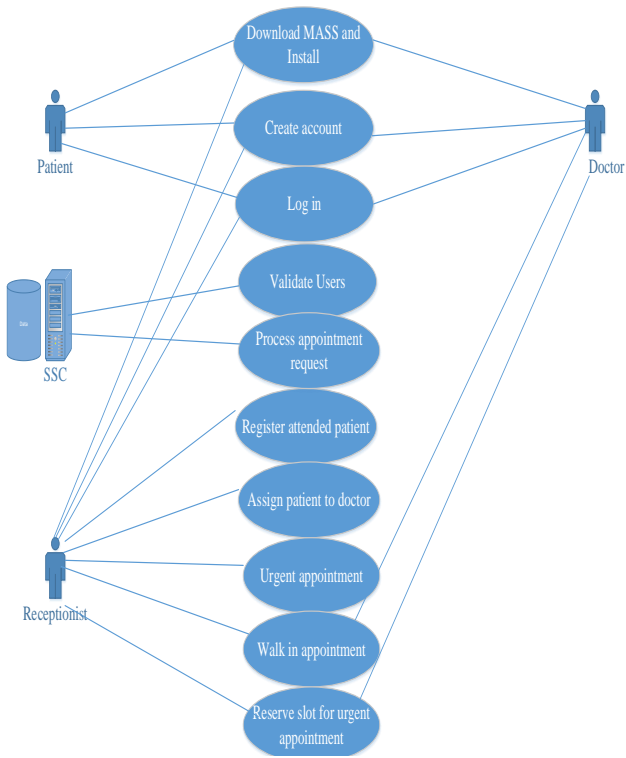

Figure.1. Involvement of actors and their collaboration among themselves and the system

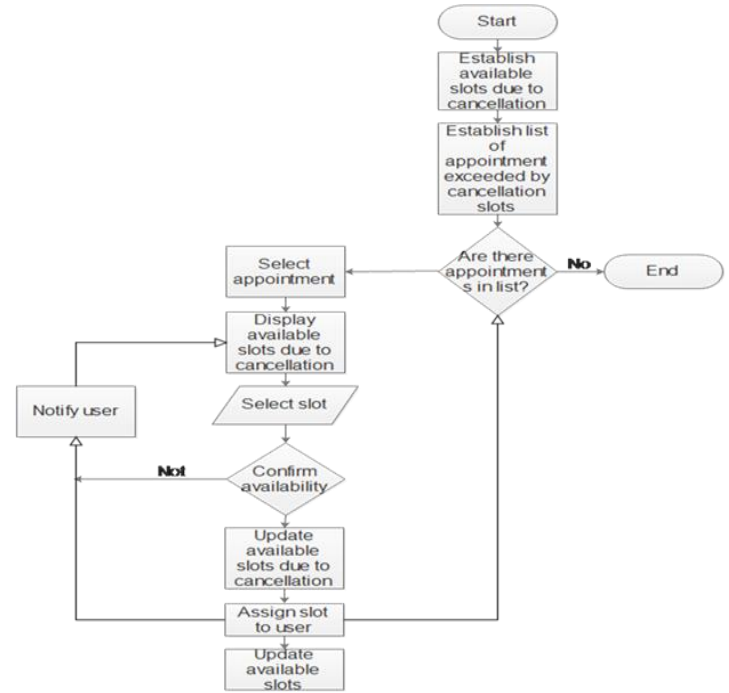

Figure 2. Process flow for slot availability due to cancellation

### B. MASS features

The proposed system will have a mechanism to display a list of available specialists and available slots, and provide notification of cancellation and postponed slots. The system will also have a mechanism for health tracking by monitoring patient's performance when visiting another hospital through retrieving the patient information from the database using the mobile phone. MASS can be used as an effective communication channel between the hospital and patients by communicating before the patient goes to the hospital. This system aims at helping patients by having appointment reminders and tracking appointments. For example, pregnant women will be provided with pregnancy tips from first week up to the last weeks of pregnancy with reminders for every appointment. After pregnancy, delivery, tips for clinic attendance will be provided through mobile phones, including reminders for attending the clinic as scheduled. Fig.2 shows the process flow for slot availability due to cancellation.

In your paper title, if the words "that uses" can accurately replace the word "using", capitalize the "u"; if not, keep using lower-cased.

## IV. DESIGNING MASS

### A. Architecture of proposed mobile appointment scheduling system

MASS is designed into two panels: (1) patients, and (2) doctors (see Fig.3). At the initial stage, users need to download MASS and install it in their devices, create an account by signing up and receive a username and password for login. And once they log in, a welcome page will display a list of offered services and patients need to select any service on the list.

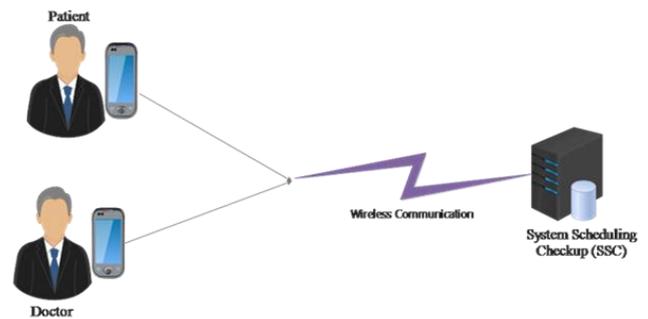

Figure. 3. Architecture of MASS

The system scheduling checkup (SSC) is the center for patient scheduling and handling; it receives and processes all appointment requests. The main task of the SSC is to receive patient requests, process and assign the patient to a doctor.

### B. Process flowchart after patient register successfully

In order to enhance appointment scheduling, MASS displays a list of medical specialists and the patient need to select a desired specialist. Once the specialist is selected, the system will establish and display available time slots and the patient is required to select the available slots. The patient has to confirm once a slot is selected so that the system can assign the time slot to the patient and update the available slots and remove the selected slot. If the selected time slot is not confirmed, the system will notify other patients about the availability of the time slot as shown in the process flow chart for appointment scheduling in Fig.4.

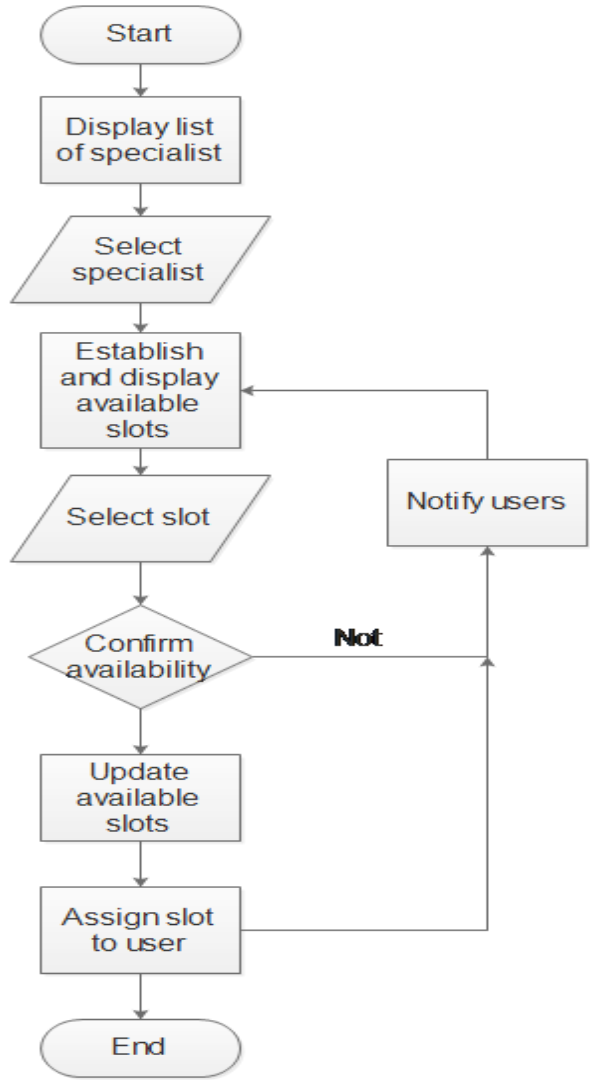

Figure.4 Process flow for appointment scheduling

During establishing and displaying available time slots, the system will display the doctor's name, schedule and timing, and it will also include detailed information of timeslot like date, month, year, time and duration to be attended. Fig.5 summarizes the data flow of information during establishing and displaying available slots, and Fig.6 when users select a medical specialist from the list of available specialists.

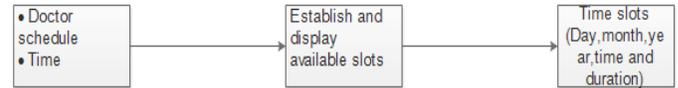

Figure.5. Data flow during establishing and displaying available slots

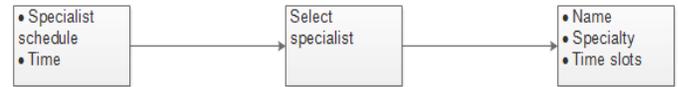

Figure.6. Data flow during functioning of selecting specialist

The overall system structure can be summarized by an algorithm shown in below.

|  | Procedure |
|---|---|
| Step 1 | Users download MASS and install in their devices |
| Step 2 | Users create account by sign up |
| Step 3 | Log in using username and password |
| Step 4 | Validate users |
| Step 5 | Display list of specialist |
| Step 6 | Select specialist on the list |
| Step 7 | Establish and display available list |
| Step 8 | Select slot |
| Step 9 | Confirm availability |
| Step 10 | If not confirmed then notify users availability of slot |
| Step 11 | If confirmed then update available slots |
| Step 12 | Assign slot to user and notify others |
| Step 13 | End procedure |

Table 1: Algorithm summary for overall system structure

## V. DEVELOPMENT OF MASS AND RESULTS

In this section, we present the results for appointment scheduling using the mobile appointment scheduling system. Initially, the user needs to register in the system by signing up to the MASS and provide username and password, which will be required during the login stage. Before making an appointment, the user is required to log in by providing username and password used during Sign-up and the system will validate the user's credentials.

### A. Appointment by day

After successful log in, the system will display a welcome page with three functionalities and the user is required to select any displayed functionality. When the user wants to make an appointment and he/she knows the name of the doctor, the user can select an appointment by day and the system will display the list of doctors available and the user will be required to select any doctor that he/she wishes to see. Once the doctor is selected, the system will retrieve detailed information about the doctor's schedule which includes status, availability, time availability for an appointment, and the doctor's specialty as shown in Fig.7 and Fig.8. Thereafter, the user will need to select an appointed day, and the system will establish available free time slots. The user is then required to select any displayed slot from the dashboard and confirm the time slot by setting the appointment. The system updates the established time slot by removing the confirmed slot from other users.

After successful log in, the system will display a welcome page with three functionalities and the user is required to select any displayed functionality. When the user wants to make an appointment and he/she knows the name of the doctor, the user can select an appointment by day and the system will display the list of doctors available and the user will be required to select any doctor that he/she wishes to see. Once the doctor is selected, the system will retrieve detailed information about the doctor's schedule which includes status, availability, time availability for an appointment, and the doctor's specialty as shown in Fig.7 and Fig.8. Thereafter, the user will need to select an appointment day, and the system will establish available free time slots. The user is then required to select any displayed slot from the dashboard and confirm the time slot by setting the appointment. The system updates the established time slot by removing the confirmed slot from other users.

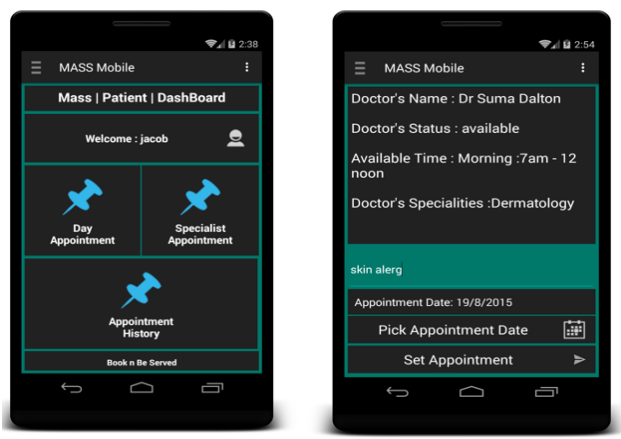

Figure.7. Welcome page   Figure.8. Set appointment Screen

### B. Appointment by Specialty

When the user wants to make an appointment by choosing a medical specialist, he/she will be required to select the appointment and the system will establish a list of available medical specialists. Each medical specialty contains a list of available doctors. The user's selected desired doctor is displayed including the doctor's schedule. The user is then required to select a time slot from the available free slots established by the system, and confirm the slot so that the system can remove it from the available established slots as shown on Fig.9 and Fig.10.

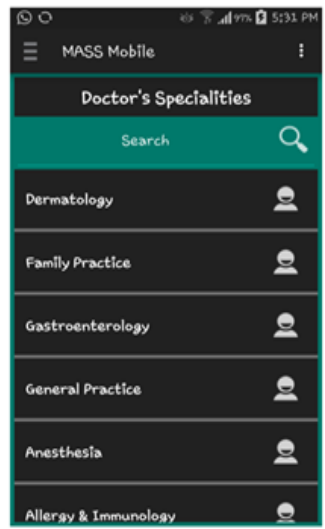 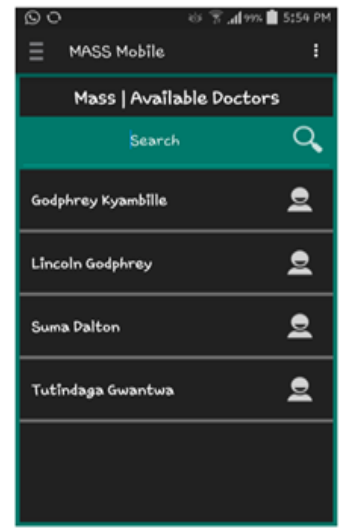

Figure.9. Doctors specialties       Figure.10. List of doctors

### C. Appointment history

The mobile application scheduling system has a mechanism that allows doctors to retrieve patient history whenever the patient visits another clinic different from the previous one in order to understand the patient's response to previous treatment before providing him/her with any medical consultation. The system has a mechanism to retrieve and systematically organize the patient history/performance of the database. Traditionally, patient's medical history is attached to the hospital clinic card. When the patient visits another clinic, the clinic card is shown to a doctor in order to understand last the appointment history. For example, when a pregnant woman visits another clinic for monthly appointment, the system will have a mechanism to retrieve last appointment detailed information as it appears on the hospital clinic card.

## VI. DISCUSSION

The design and development of the mobile appointment scheduling system was done using MYSQL with WAMP server and PHP. The database system is developed with MySQL which is an open source application possessed and overseen by Sun Microsystems and gained by the Oracle Corporation. The scripting was done by utilizing PHP.

Let us consider an appointment scheduling scenario involving a patient cancelling an appointment. Once the patient is assigned a time slot successfully and decides to cancel the appointment, MASS will have a mechanism for notifying other patients on the availability of a slot for any patient that needs to reschedule an appointment. The system will establish the available time slots due to cancellation and display the updated available slots. Any patient making an appointment can select the available slots displayed and confirm to schedule the appointment so that the slot can be removed from the list. If the patient fails to confirm the appointment, the system will display the time slot to other patients as a free slot. In this scenario, MASS will enhance appointment scheduling by informing patients whenever there is free slot due to cancellation and patients whose appointment is deactivated by cancellation will be required either to select the available free slot due to cancellation or to remain in the existing timeslot.

Another scenario is for doctors postponing appointments with patients. Once introduced in hospitals, MASS will improve appointment scheduling by establishing the available time slots after a doctor postpones appointment. Postponing the appointments may create time slots which can other patients can use to see the doctor. In case of a doctor postponing appointments and therefore creating time slots to engage other patients, the system will check if there are appointment requests in the list. If pending appointment requests are found, then the appointments will be selected and a list of available time slots resulting from a doctor's appointment postponement will be displayed. The user will be required to select a time slot from the available slots and confirm to take the slot. Once confirmed, the system will update the available slots found due to the doctor's appointment postponement, and the selected slot will be assigned to the user. To avoid other users to choose the selected time slot, the system will notify other users by updating the available slots and hide the selected slot. If the user does not confirm the selected time slot, then the system will notify other users of the availability of the slot (see Fig.11).

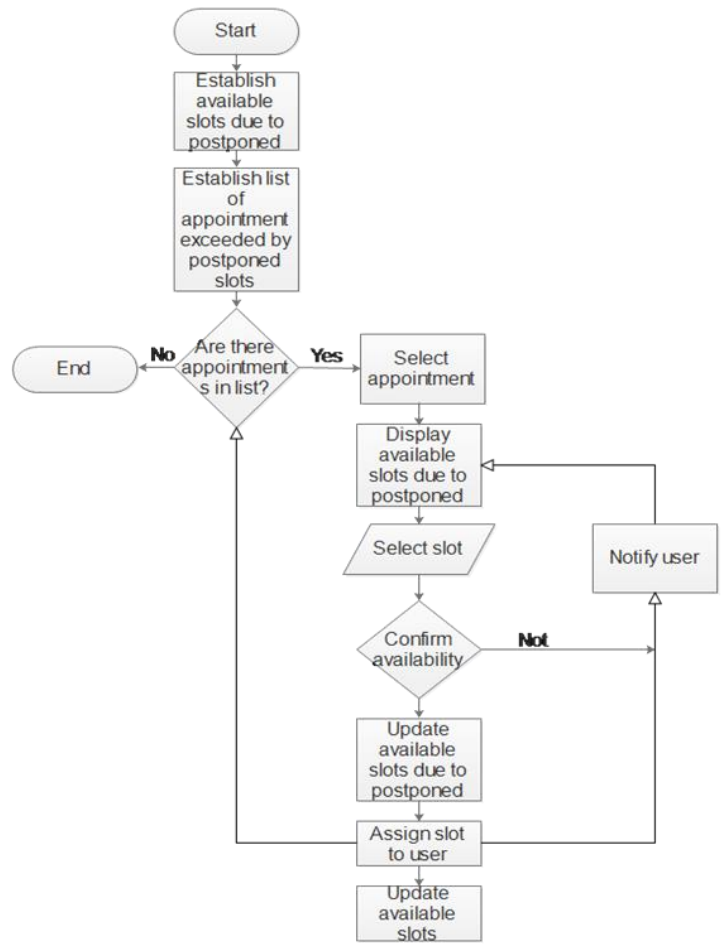

Figure.11. Process flow for doctor postponed appointment

## VII. CONCLUSION AND FUTURE WORK

Scheduling appointments appropriately and resourcefully is vital to the smooth process of the hospitals' service delivery. Working in the new era of science and technology, people have the slight patience for waiting in the queue at the hospitals. Patients, who make appointments weeks in advance, want to be attended within 20 minutes after showing up at the hospital. They prefer to be given a specific time for seeing the doctor rather than arriving at the hospital and wait for an open moment. Doctors need a smooth tide of patients when attending the scheduled patients. Conferring the challenges facing existing patient appointment systems, we are proposing an integrated mobile appointment scheduling system that will enhance appointment scheduling in hospitals with the aim of simplifying patients and doctors' task. In our system, the SSC gathers information from the users and schedule patients based on the availability of doctor time slots. In employing the proposed system, patients will be more relaxed whenever they make appointments without standing in the long queue as the system would replicate tasks which would otherwise be carried out hospital personnel and patients. Doctors will be more comfortable in attending patients in a systematic flow as the system manages the appointment requests and scheduling. In future, the system can be developed to direct appointment requests to another hospital where doctors with similar medical expertise are working. Moreover, providing automatic calls as reminders when the appointed day approaches or arrives is a vital feature of the system.

## AUTHORS PROFILE

Godphrey Kyambille is a Tutorial Assistant in the Computer Engineering department at Mbeya University of Science and Technology. Currently he is studying Master's degree in Information and Communication Science and Engineering, specializing in Information Technology System Development and Management at Nelson Mandela African Institution of Science and Technology.He currently lives in Tanzania

Khamisi Kalegele is a lecture at Nelson Mandela African Institution of Science and Technology school of Computational and Communication Sciences and Engineering (CoCSE). He currently lives in Tanzania.